\documentclass[a4paper,floatfix,amsmath,amssymb,fleqn,prb,twoside,showkeys,showpacs]{revtex4}
\usepackage{graphicx} 
\usepackage{epsf}
\usepackage{dcolumn}
\usepackage{bm}
\usepackage{color}
\begin{document}
\newcommand{\newc}{\newcommand}
\newc{\mbf}{\mathbf}
\newc{\boma}{\boldmath}
\newc{\beq}{\begin{equation}}
\newc{\eeq}{\end{equation}}
\newc{\beqar}{\begin{eqnarray}}
\newc{\eeqar}{\end{eqnarray}}
\newc{\beqa}{\begin{eqnarray*}}
\newc{\eeqa}{\end{eqnarray*}}
\newc{\bd}{\begin{displaymath}}
\newc{\ed}{\end{displaymath}}

\title{Understanding d'Alembert's principle: System of Pendulums}

\author{Subhankar Ray}
\email{sray_ju@rediffmail.com, sray@phys.jdvu.ac.in (S. Ray)}
\affiliation{Department of Physics, Jadavpur University, Calcutta 700 032, India}
\author{J. Shamanna}
\email{jshamanna@rediffmail.com (J. Shamanna)}
\affiliation{Department of Physics, University of Calcutta, Calcutta 700 009, India}

\date{May 26, 2006}

\begin{abstract}
Lagrangian mechanics uses d'Alembert's principle of 
zero virtual work as an important starting point.
The orthogonality of the force of constraint and virtual 
displacement is emphasized in literature, without a clear 
warning that this is true usually for a single particle system.
For a system of particles connected by constraints,
it is shown, that the virtual work of the entire system is zero,
even though the virtual displacements of the particles
are not perpendicular to the respective constraint forces.
It is also demonstrated why d'Alembert's principle involves
virtual work rather than the work done by constraint forces
on allowed displacements.
\end{abstract}
\maketitle

\section{Introduction}
The principle of zero work by constraint forces on virtual displacement,
also known as d'Alembert's principle,
is an important step in formulating and solving a mechanical problem with constraints
\cite{goldstein,greenwood,hylleraas,sommer,taylor}. 
In the simple systems widely used in literature, e.g., a single 
particle rolling down a frictionless incline, or a simple pendulum with
inextensible ideal string, the force of constraint is perpendicular to
the virtual displacement. 
This results in zero virtual work by constraint forces. 
It is often tempting to assume that the 
constraint forces are always orthogonal to 
respective virtual displacements, even for a 
system of particles.
d'Alembert's principle then seems to be a consequence of 
this orthogonality \cite{goldstein,hylleraas,taylor}.

In this article we study two simple systems: a double pendulum and an $N$-pendulum.
In these systems it is observed that, the virtual displacements
are not perpendicular to the respective constraint forces acting on 
individual particles (pendulum bobs).
However, d'Alembert's principle of zero virtual work still holds for the
systems as a whole. 
In these problems, the principle of zero virtual work is a consequence
of, (i) the relation between the virtual displacements of
coupled components (neighbouring bobs), and (ii) the appearance of
(Newtonian) action-reaction pairs in forces of constraint between
neighbouring particles.
Thus d'Alembert's principle is more subtle and involved than it is often thought to be.
Greenwood has rightly said, 
``... workless constraints do no work on the system as a whole 
in an arbitrary virtual displacement. Quite possibly, 
however, the workless constraint forces 
will do work on individual particles of the system'' \cite{greenwood}.

\section{Double Pendulum}
Let us first consider a double pendulum, with inextensible ideal strings 
of lengths $L_1$ and $L_2$.
Let $\mbf{r}_1$ and $\mbf{r}_2$ denote the instantaneous
positions of the pendulum bobs $\mathcal{P}_1$ and $\mathcal{P}_2$,
with respect to the point of suspension (see figure 1). 
When the system is suspended 
from a stationary support; the holonomic, scleronomous constraint equations are,
\beq\label{2const}
| \mbf{r}_1 | = L_1, \hskip 1cm 
|\mbf{r}_1 - \mbf{r}_{2}| = L_2.
\eeq
The equations for allowed velocities are obtained by differentiating
(\ref{2const}) with respect to $t$,
\beq\label{2avel}
\mbf{r}_1 \cdot \mbf{v}_1 =0, \hskip 1cm
(\mbf{r}_2-\mbf{r}_1) \cdot (\mbf{v}_2-\mbf{v}_1)=0.
\eeq
Thus, the allowed velocity $\mbf{v}_1$ of $\mathcal{P}_1$ is orthogonal 
to its position vector $\mbf{r}_1$.
The relative velocity of the second bob $(\mbf{v}_2-\mbf{v}_1)$ 
with respect to the first, is orthogonal to the
relative position vector $\mbf{r}_{21}=(\mbf{r}_2-\mbf{r}_1)$.
Let $\hat{\mbf{n}}_{1}$ and $\hat{\mbf{n}}_{2}$ be unit vectors 
orthogonal to ($\mbf{r}_1$) and ($ \mbf{r}_2-\mbf{r}_1$)
respectively,
\beq\label{2unit}
\hat{\mbf{n}}_{1} \cdot \mbf{r}_1 =0 \hskip 1cm \mbox{and}
\hskip 1cm \hat{\mbf{n}}_{2} \cdot (\mbf{r}_2-\mbf{r}_1)=0.
\eeq
From (\ref{2avel}) and (\ref{2unit}) we obtain the allowed velocities as,
\beq\label{2v1v2}
\mbf{v}_1 =b_1 \hat{\mbf{n}}_{1},
\hskip 1cm \mbf{v}_{2}-\mbf{v}_{1} =  b_2 \hat{\mbf{n}}_{2}
\hskip 0.5cm \Rightarrow 
\hskip 0.5cm \mbf{v}_{2}=\mbf{v}_{1} + b_2 \hat{\mbf{n}}_{2} 
\eeq
where $b_1$ and $b_2$ are arbitrary real constants, denoting the magnitude
of the relevant vectors.

\begin{figure}[h]
{\includegraphics{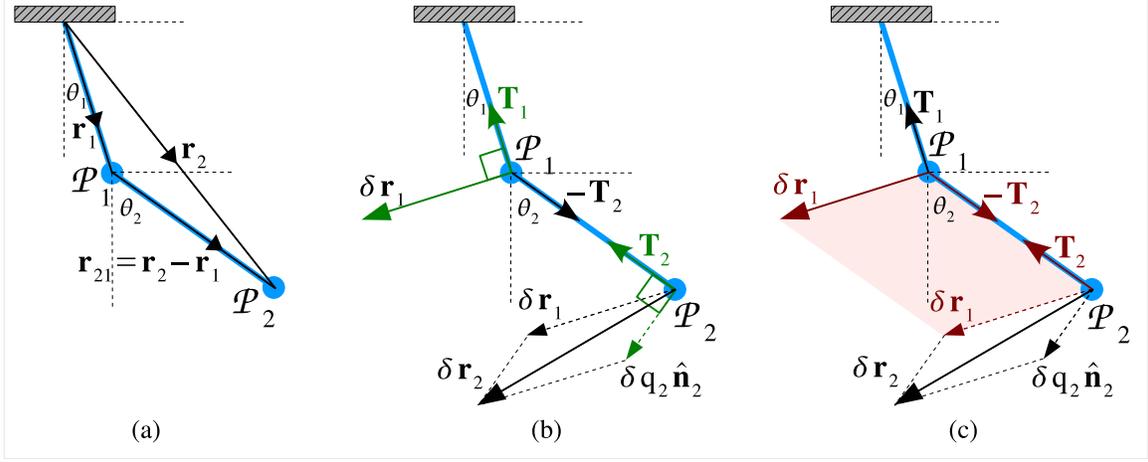}}
\caption{Double pendulum: (a) position vectors, (b) orthogonality of tensions
to part of virtual displacements, (c) cancellation of part of virtual work
related to action-reaction pair (tension)}
\label{dpend_stat}
\end{figure}
The virtual velocities are defined as a difference between two allowed 
velocities, $\widetilde{\mbf{v}}_k=\mbf{v}_k -\mbf{v}_k'$ \cite{sj_ejp_27}.
The allowed displacements $d \mbf{r}_k$, and the virtual
displacements $\delta \mbf{r}_k$
are then given by,
\beqar\label{2disp}
&& d \mbf{r}_1= \mbf{v}_{1} \, dt = dq_1 \hat{\mbf{n}}_1, \nonumber \\
&& \delta \mbf{r}_1= d \mbf{r}_1 - d \mbf{r}_1'
=(dq_1  -dq_1') \hat{\mbf{n}}_1
= \delta q_1 \hat{\mbf{n}}_{1}. 
\eeqar
\beqar\label{3disp}
&& d \mbf{r}_2= \mbf{v}_2 dt = \mbf{v}_1 dt + b_2 \hat{\mbf{n}}_2 dt 
= d \mbf{r}_1 + d q_2 \hat{\mbf{n}}_2, \nonumber \\
&& 
\delta \mbf{r}_2 = (d \mbf{r}_1 + dq_2 \hat{\mbf{n}}_2) - (d \mbf{r}_1'+
+ dq_2' \hat{\mbf{n}}_2)
= \delta \mbf{r}_1 +\delta q_2 \hat{\mbf{n}}_{2}.
\eeqar
It may be noted that, under the given holonomic, scleronomous constraints
(velocity and time independent),
the sets of allowed and virtual displacements are equivalent.

A set of allowed displacements $\{d \mbf{r}_1, d \mbf{r}_2 \}$
is obtained by a specific choice of the numbers $\{b_1,b_2 \}$ or 
$\{ d q_1, d q_2 \}$. By making different choices of the set $\{ d q_1, d q_2 \}$
we get a whole family of allowed displacements, $\{ d \mbf{r}_k \}$, where,
\bd
\{ d \mbf{r}_k \}  \Rightarrow \{d \mbf{r}_1, d \mbf{r}_2 \}_{\{b_1, b_2\}} = 
\{d \mbf{r}_1, d \mbf{r}_2 \}_{\{d q_1, d q_2\}}.
\ed
Or more precisely,
\bd
\{ d \mbf{r}_k \}  \Rightarrow \{ \{ 
d \mbf{r}_1, d \mbf{r}_2 \}_{\{b_1, b_2\}} | b_1, b_2 \in
\Re \}
\ed
where $\Re$ is the set of real numbers.
Similarly, by choosing different set of quantities $\{ \delta q_1, \delta q_2 \}$
we get the family of virtual displacements, $\{ \delta \mbf{r}_k \}$, where,
\bd
\{ \delta \mbf{r}_k \}  \Rightarrow \{\delta \mbf{r}_1, \delta \mbf{r}_2 \}
_{\{\delta q_1, \delta q_2\}}.
\ed
Thus, it is easy to see that for holonomic, scleronomous constraints,
the set of all possible allowed displacements is the same as the set of 
all possible virtual displacements. This is in agreement with the fact that for
holonomic, scleronomous systems, the sets $\{ d \mbf{r}_k \}$ and 
$\{ \delta \mbf{r}_k \}$ satisfy the same equations\cite{sj_ejp_eqn}, namely,
\bd
\begin{array}{l l}
\mbf{r}_1 \cdot d \mbf{r}_1 = 0, \hspace*{1cm} & (\mbf{r}_2 - \mbf{r}_1) \cdot ( d\mbf{r}_2 - d\mbf{r}_1 ) =0 \\
\mbf{r}_1 \cdot \delta \mbf{r}_1 = 0, & (\mbf{r}_2 - \mbf{r}_1) \cdot ( \delta \mbf{r}_2 - \,\delta \mbf{r}_1 ) =0 \\
\end{array}
\ed
As the pendulums are suspended by inextensible ideal strings, one may assume
that the tensions in the strings act along their lengths. 
This essentially implies that there is no shear in the string to transmit 
transverse force. Thus the tension
$\mbf{T}_1$ is along $(-\mbf{r}_1)$ and tension $\mbf{T}_2$ is along
$(\mbf{r}_1-\mbf{r}_2)$.
Hence, the virtual displacement $\delta \mbf{r}_1$ for the first pendulum
is perpendicular to the tension $\mbf{T}_1$, but the virtual displacement 
$\delta \mbf{r}_2$ of the second pendulum is not perpendicular to $\mbf{T}_2$.
\beqar\label{t2dr2}
\mbf{T}_1 \cdot  \delta \mbf{r}_1 &=& 0 \nonumber \\ 
\mbf{T}_2 \cdot  \delta \mbf{r}_2 &=& \mbf{T}_2 \cdot (\delta \mbf{r}_1 + \delta q_2 
{\hat{\mbf{n}}}_2 ) =  \mbf{T}_2 \cdot \delta \mbf{r}_1 + (\mbf{T}_2 \cdot \hat{\mbf{n}}_2)
\delta q_2 = \mbf{T}_2 \cdot \delta \mbf{r}_1
\eeqar
At this stage one may appreciate that $\mbf{T}_1$ is not the entire
force of constraint on $\mathcal{P}_1$. 
As a reaction to $\mathcal{P}_1$ pulling $\mathcal{P}_2$ with a tension
$\mbf{T}_2$, the second bob $\mathcal{P}_2$ pulls the first bob $\mathcal{P}_1$
with a tension $( - \mbf{T}_2)$. Thus the virtual work by constraint forces
acting on $\mathcal{P}_1$ and $\mathcal{P}_2$ are given by,
\beqar \label{w1w2}
&&\delta \mathcal{W}_1 = \mbf{R}_1 \cdot \delta \mbf{r}_1 = (\mbf{T}_1 - \mbf{T}_2) \cdot  \delta \mbf{r}_1 = - \mbf{T}_2  \cdot  \delta \mbf{r}_1 \neq 0  \nonumber \\
&&\delta \mathcal{W}_2 = \mbf{R}_2 \cdot \delta \mbf{r}_2 =  \mbf{T}_2 \cdot  (\delta \mbf{r}_1
+\delta q_2 \hat{\mbf{n}}_2) =\mbf{T}_2 \cdot \delta \mbf{r}_1 \neq 0. 
\eeqar
Although neither $\delta \mathcal{W}_1$ nor $\delta \mathcal{W}_2$ is zero, their sum
adds up to zero. 
Thus the ``equal and opposite'' Newtonian reaction comes to our rescue, and 
we have a cancellation in the total virtual work.
\beq\label{wsum}
\delta \mathcal{W}_1 + \delta \mathcal{W}_2 =
\mbf{R}_1 \cdot \delta \mbf{r}_1 + \mbf{R}_2 \cdot \delta \mbf{r}_2 = 
 - \mbf{T}_2  \cdot  \delta \mbf{r}_1 + \mbf{T}_2 \cdot \delta \mbf{r}_1 =0
\eeq
This shows that in the case of a double pendulum with stationary support,
d'Alembert's principle utilizes the equal and opposite nature of the
tensions between neighbouring bobs (Newtonian action-reaction pair).

Let us now consider the double pendulum with a moving point of suspension.
This gives us a system with a rheonomous constraint.
Let the velocity of the point of suspension be $\mbf{v}_0$.
The constraint equations in this case are,
\beq
|\mbf{r}_1 - \mbf{v}_0 t| = L_1, \hskip 1cm |\mbf{r}_2 - \mbf{r}_{1}|=L_2
\eeq
The only non-trivial modification is that, the virtual displacements are no longer
equivalent to the allowed displacements. 

For the first bob $\mathcal{P}_1$,
\beqar\label{mv_disp1}
&& d \mbf{r}_1= \mbf{v}_{1} \, dt = \mbf{v}_{0} \, dt + dq_1 \hat{\mbf{n}}_1, \nonumber \\
&& \delta \mbf{r}_1= d \mbf{r}_1 - d \mbf{r}_1' 
=( \mbf{v}_{0} dt \hskip -.6cm \backslash \,\,\, \backslash \, + dq_1 \hat{\mbf{n}}_1 ) 
-( \mbf{v}_{0} dt \hskip -.6cm \backslash \,\,\, \backslash \, + dq_1' \hat{\mbf{n}}_1)
=( dq_1 -dq_1') \hat{\mbf{n}}_1
= \delta q_1 \hat{\mbf{n}}_{1}. 
\eeqar
Therefore virtual displacement $\delta \mbf{r}_1$ is a vector along $\hat{\mbf{n}}_1$, whereas 
allowed displacement $d\mbf{r}_1$ is sum of a vector along  $\hat{\mbf{n}}_1$ and a vector 
along $\mbf{v}_{0}$.
For the second bob $\mathcal{P}_2$,
\beqar\label{mv_disp2}
&& d \mbf{r}_2= \mbf{v}_2 dt = \mbf{v}_1 dt + b_2 \hat{\mbf{n}}_2 dt 
= \mbf{v}_0 dt + d q_1 \hat{\mbf{n}}_1 + d q_2 \hat{\mbf{n}}_2 \nonumber \\
&& \delta \mbf{r}_2
= (\mbf{v}_0 dt \hskip -.6cm \backslash \,\,\, \backslash \, 
+ d q_1 \hat{\mbf{n}}_1 + d q_2 \hat{\mbf{n}}_2) - 
     (\mbf{v}_0 dt \hskip -.6cm \backslash \,\,\, \backslash \,
+ d q_1' \hat{\mbf{n}}_1 + d q_2' \hat{\mbf{n}}_2)
= \delta q_1 \hat{\mbf{n}}_{1} +\delta q_2 \hat{\mbf{n}}_{2}.
\eeqar
Thus $d \mbf{r}_1$ and $d \mbf{r}_2$ are not equivalent to
$\delta \mbf{r}_1$ and $\delta \mbf{r}_2$.  
However the relation between $ \delta \mbf{r}_1$ and $\delta \mbf{r}_2$
remains the same as in the case of a double pendulum with stationary support.
\beq\label{mv_vd}
\delta \mbf{r}_2 = \delta \mbf{r}_1 + \delta q_2 \hat{\mbf{n}}_{2}
\eeq
Hence the above inferences, in particular, (\ref{t2dr2}), (\ref{w1w2}), (\ref{wsum}) are 
true even in this case.

\section{\textbf{\textit{N}}-Pendulum}
\begin{figure}[h]
\resizebox{!}{!}
{\includegraphics{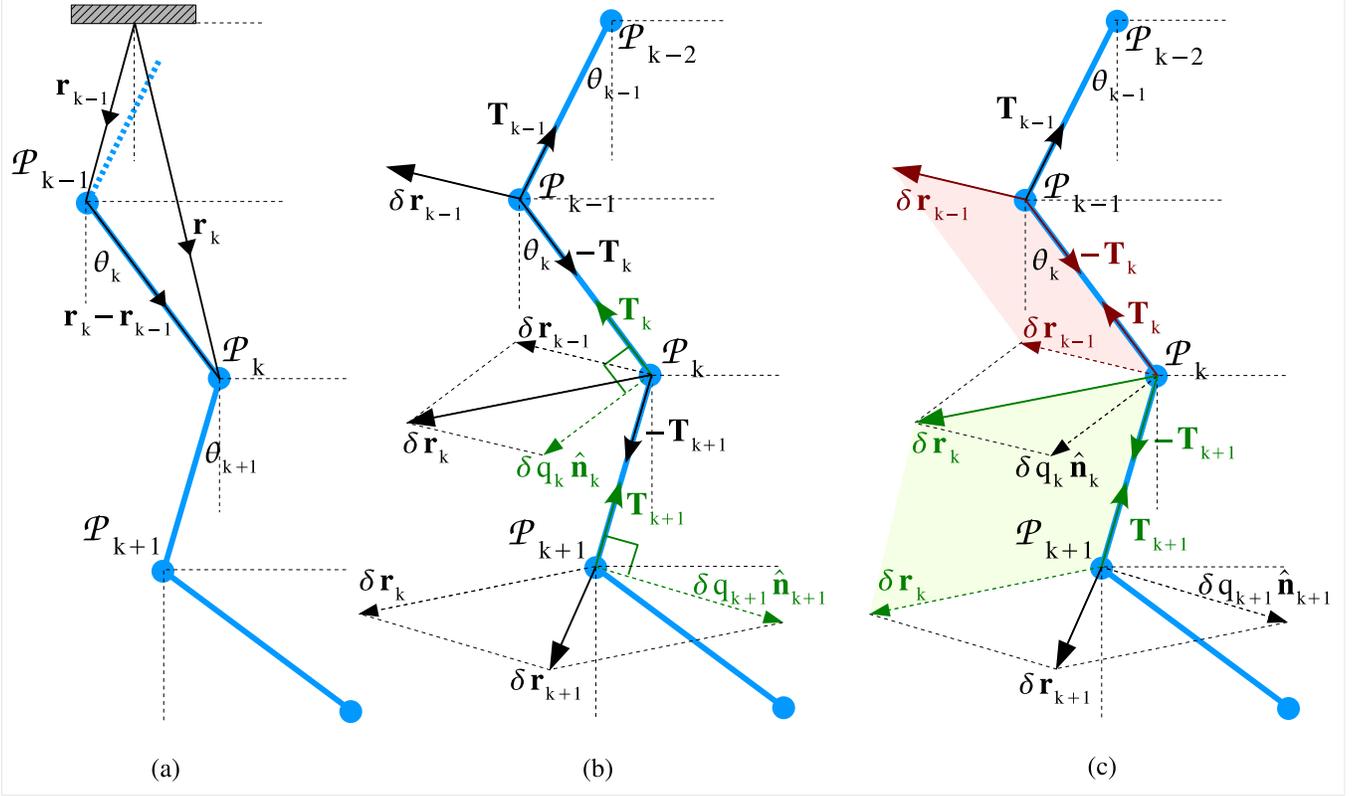}}
\caption{$N$-pendulum: (a) position vectors (position of the point of suspension in relation to the pendulum is schematic), (b) orthogonality of tensions
to part of virtual displacements, (c) cancellation of part of virtual work
related to action-reaction pair (tension)}
\label{npend}
\end{figure}

It is instructive to repeat the above exercise for a system of $N$-pendulum 
joined end to end by inextensible, ideal strings.
Let $\mbf{r}_1, \mbf{r}_2,\dots, \mbf{r}_N$ denote the instantaneous 
position vectors of pendulum bobs $\mathcal{P}_1, \mathcal{P}_2,\dots, \mathcal{P}_N$ 
respectively, as shown in figure \ref{npend}. 
The constraint equations for this system are,
\beq\label{nconst}
|\mbf{r}_1 | = L_1, \hskip 1cm
|\mbf{r}_k-\mbf{r}_{k-1}| = L_k, \hskip 2cm  k=2,3,\dots,N. 
\eeq
The equations for allowed velocities, obtained by differentiating the above
equations, are
\beq\label{nvel}
\mbf{r}_1 \cdot \mbf{v}_1 = 0, \hskip 1cm
(\mbf{r}_k-\mbf{r}_{k-1}) \cdot (\mbf{v}_k-\mbf{v}_{k-1})=0, 
\hskip 2cm k=2,3,\dots,N.
\eeq
Let us introduce unit vectors 
$\hat{\mbf{n}}_1, \hat{\mbf{n}}_2, \dots \hat{\mbf{n}}_N$,
where $\hat{\mbf{n}}_k$ is 
normal to the relative position of the bob $\mathcal{P}_k$
with respect to $\mathcal{P}_{k-1}$, i.e.,
$(\mbf{r}_k-\mbf{r}_{k-1})$.
\beq\label{nunit}
\hat{\mbf{n}}_1 \cdot \mbf{r}_1 = 0, \hskip 1cm
\hat{\mbf{n}}_k \cdot (\mbf{r}_k-\mbf{r}_{k-1})=0, 
\hskip 2cm k=2,3,\dots, N.
\eeq
From (\ref{nvel}) and (\ref{nunit}), the allowed velocities are given by
\beq\label{nv1v2}
\mbf{v}_1 = b_1 \hat{\mbf{n}}_1, \hskip 1cm 
\mbf{v}_k-\mbf{v}_{k-1}=b_k \hat{\mbf{n}}_k
\hskip 0.5cm \Rightarrow \hskip 0.5cm 
\mbf{v}_k = \mbf{v}_{k-1} + b_k \hat{\mbf{n}}_k, \hskip 2cm k=2,3,\dots, N. 
\eeq
where $\{ b_k, k = 1, \dots, N \}$ are a set of real constants
denoting the magnitude of the relevant vectors.
The allowed displacements $d\mbf{r}_k$ and the virtual displacements
$\delta \mbf{r}_k$ are,
\beq\label{ndisp}
\begin{array}{l l}
d\mbf{r}_1=b_1 \hat{\mbf{n}}_1 dt = dq_1 \hat{\mbf{n}}_1,
       & \delta \mbf{r}_1= \delta q_1 \hat{\mbf{n}}_1, \nonumber \\
d\mbf{r}_2=d\mbf{r}_1+b_2 \hat{\mbf{n}}_2 dt = d\mbf{r}_1+dq_2 \hat{\mbf{n}}_2,
       & \delta \mbf{r}_2= \delta \mbf{r}_1+\delta q_2 \hat{\mbf{n}}_2  \nonumber \\
\vdots & \vdots \nonumber \\
d\mbf{r}_N=d\mbf{r}_{N-1}+b_N \hat{\mbf{n}}_N dt = d\mbf{r}_{N-1}+dq_N \hat{\mbf{n}}_N, \hspace*{1cm}
       & \delta \mbf{r}_N=\delta \mbf{r}_{N-1}+\delta q_N \hat{\mbf{n}}_N 
\end{array}
\eeq
As is noted in the previous section for case of double pendulum, 
due to the holonomic, scleronomous
nature of the constraints, the sets of allowed and virtual displacements are 
equivalent.
From the above equations one can see that the virtual displacement
of each pendulum ($\delta \mbf{r}_k$) is a vector sum of the virtual displacement
of the previous pendulum ($\delta \mbf{r}_{k - 1}$) and a component along the 
unit normal $\hat{\mbf{n}}_k$. 
Hence the virtual displacements (with the exception of $\delta \mbf{r}_1$) are not
orthogonal to the corresponding relative position vectors.

Let us now consider the constraint forces on each individual pendulum 
bob $\mathcal{P}_k$.
The bob $\mathcal{P}_k$ is pulled towards its point of suspension (the previous bob
$\mathcal{P}_{k-1}$) by a tension $\mbf{T}_k$ along $(\mbf{r}_{k-1}-\mbf{r}_k)$. 
The next pendulum bob, $\mathcal{P}_{k+1}$, is pulled towards $\mathcal{P}_k$ 
by a tension $\mbf{T}_{k+1}$ along $(\mbf{r}_k-\mbf{r}_{k+1})$. 
In response to this, a reaction force $(-\mbf{T}_{k+1})$ acts on the bob $\mathcal{P}_k$ 
along $(\mbf{r}_{k+1}-\mbf{r}_k)$.
Thus between any two neighbouring pendulum bobs, there exists a pair of 
{\it equal and opposite} action-reaction forces.
The total force on $\mathcal{P}_k$ is $(\mbf{T}_k -\mbf{T}_{k+1})$ for $k = 1,2,\dots,N-1$.
However, for the last pendulum $\mathcal{P}_N$, the net constraint force is $\mbf{T}_N$.

The virtual work done by the constraint forces at different system points (particle
positions) are,
\beqar\label{nw1}
&& \delta \mathcal{W}_1 = \mbf{R}_1 \cdot \delta \mbf{r}_1 =
(\mbf{T}_1-\mbf{T}_{2}) \cdot \delta \mbf{r}_1 
=\mbf{T}_1 \cdot \delta \mbf{r}_1 -\mbf{T}_2 \cdot \delta \mbf{r}_1 \nonumber \\
&& \delta \mathcal{W}_k= \mbf{R}_k \cdot \delta \mbf{r}_k =
(\mbf{T}_k-\mbf{T}_{k+1}) \cdot \delta \mbf{r}_k 
=\mbf{T}_k \cdot (\delta \mbf{r}_{k-1}+\delta q_k \hat{\mbf{n}}_k)-
\mbf{T}_{k+1} \cdot \delta \mbf{r}_k \hskip .5cm k =2,\dots, N-1 \nonumber \\
&& \delta \mathcal{W}_N= \mbf{R}_N \cdot \delta \mbf{r}_N =
\mbf{T}_N \cdot (\delta \mbf{r}_{N-1}+\delta q_N \hat{\mbf{n}}_N)
\eeqar
As the strings of the pendulums are ideal, the tension $\mbf{T}_k$ acts along 
the length of the string, i.e., $(\mbf{r}_{k-1}-\mbf{r}_k)$. 
Thus the tension $\mbf{T}_k$ is normal to the unit vector $\hat{\mbf{n}}_k$. 
The above virtual work elements become,
\beq\label{nw2}
\delta \mathcal{W}_1= -\mbf{T}_{2} \cdot \delta \mbf{r}_1 ,
\hskip 1cm \delta \mathcal{W}_k= 
\mbf{T}_k \cdot \delta \mbf{r}_{k-1} -\mbf{T}_{k+1} \cdot \delta \mbf{r}_k ,
\hskip 1cm \delta \mathcal{W}_N= \mbf{T}_N \cdot\delta \mbf{r}_{N-1}.
\eeq
It is clear that the virtual work at each system point $\mathcal{P}_k$ is 
non-zero. However if we sum the virtual work at all these system points,
we observe a mutual cancellation and the total virtual work vanishes.
\beqar\label{nwsum}
{\textstyle \sum_{k=1}^{N}} \,\, \delta \mathcal{W}_k = &- \mbf{T}_2 \cdot \delta \mbf{r}_1 \! 
&+ \, (\mbf{T}_2 \cdot \delta \mbf{r}_1 - \mbf{T}_3 \cdot \delta \mbf{r}_2) 
+\dots \dots \dots \nonumber \\ 
& \hfill {\dots \dots} \! &+ \, (\mbf{T}_{j-1} \cdot \delta \mbf{r}_{j-2} - \mbf{T}_{j} \cdot \delta \mbf{r}_{j-1}) 
+ (\mbf{T}_j \cdot \delta \mbf{r}_{j-1} - \mbf{T}_{j+1} \cdot \delta \mbf{r}_j)
+ (\mbf{T}_{j+1} \cdot \delta \mbf{r}_{j} 
-\mbf{T}_{j+2} \cdot \delta \mbf{r}_{j+1}) + \dots \nonumber \\
& \hfill {\dots \dots} \! &+ \, (\mbf{T}_{N-1} \cdot \delta \mbf{r}_{N-2} - \mbf{T}_N \cdot \delta \mbf{r}_{N-1}) + \mbf{T}_N \cdot \delta \mbf{r}_{N-1} = 0
\eeqar
This vanishing of total virtual work is a consequence of
(i) definition of virtual displacement, (ii) appearance of action-reaction pairs
in the forces of constraint.
The virtual work connected to each bob $\mathcal{P}_k$, is
composed of three parts, 
(i) virtual work by the tension $\mbf{T}_{k}$ on the component of virtual displacement 
$\delta q_k \hat{\mbf{n}}_k$ orthogonal to the relative position vector,
(ii) virtual work by the tension $\mbf{T}_{k}$ on part of the virtual displacement 
$\delta \mbf{r}_{k-1}$ related to that of the previous bob, and
(iii) virtual work by the reaction tension $(- \mbf{T}_{k+1})$ 
(acting towards the next bob $\mathcal{P}_{k+1}$) 
on the virtual displacement 
$\delta \mbf{r}_{k}=\delta \mbf{r}_{k-1}+\delta q_k \hat{\mbf{n}}_k$.
The first component for each $\mathcal{P}_k$ vanishes because of orthogonality
of the related force and virtual displacement.
This is shown schematically in figure \ref{npend}(b).
Due to the ``equal and opposite'' nature of action reaction pairs,
and existence of a common term in the virtual displacement of neighbouring bobs,
the other terms for each bob cancel with the related terms of its neighbours.
Shaded areas in figure \ref{npend}(c) illustrate this cancellation.

\begin{table}[h]
\caption{Virtual work and d'Alembert's principle for simple and $N$-pendulum}
\begin{tabular}{||l|c|c|c||}
\hline 
&& \multicolumn{2}{c||}{} \\
System  &  Stationary support & \multicolumn{2}{c||}{Moving support} \\
Pendulums with fixed string length &  scleronomous & \multicolumn{2}{c||}{rheonomous} \\
holonomic constraints & ${ (\delta \mbf{r}_k \sim d \mbf{r}_k) }^{\dagger}$ & 
          \multicolumn{2}{c||}{ ${(\delta \mbf{r}_k \not\sim d \mbf{r}_k )}^{\star}$} \\
&& \multicolumn{2}{c||}{} \\
\hline
\hspace*{6.0cm} && \hspace*{3.5cm} & \hspace*{3.5cm} \\
Simple pendulum  &  $\mbf{T} \cdot \delta \mbf{r}=0$ &  $\mbf{T} \cdot d \mbf{r} \neq 0$ &
                              $\mbf{T} \cdot \delta \mbf{r} = 0$ \\
\hfill  {\scriptsize{$\mbf{R}=\mbf{T}$}} & 
   \framebox[1.8cm]{$\mbf{R} \cdot \delta \mbf{r}=0$} &  
                    $\mbf{R} \cdot d \mbf{r} \neq 0$ &
   \framebox[1.8cm]{$\mbf{R} \cdot \delta \mbf{r} = 0$} \\
&&& \\
\hline
&&&\\
$N$-pendulum & $\mbf{T}_k \cdot  \delta \mbf{r}_k \neq 0$ & 
                  $\mbf{T}_k \cdot  d \mbf{r}_k \neq 0$  &
                  $\mbf{T}_k \cdot  \delta \mbf{r}_k \neq 0$ \\
& {\scriptsize{$\displaystyle{\sum_{k=1}^{N}}$}} $\,\mbf{T}_k \cdot \delta \mbf{r}_k \neq 0$ &
   {\scriptsize{$\displaystyle{\sum_{k=1}^{N}}$}} $\,\mbf{T}_k \cdot d \mbf{r}_k \neq 0$ &
   {\scriptsize{$\displaystyle{\sum_{k=1}^{N}}$}} $\,\mbf{T}_k \cdot \delta \mbf{r}_k \neq 0$ \\
\cline{2-4}
&&&\\
\hfill {\scriptsize{$\mbf{R}_k = (\mbf{T}_k-\mbf{T}_{k+1}), \,\,k=1,...N-1$}} &
                  $\mbf{R}_k \cdot  \delta \mbf{r}_k \neq 0$ & 
                  $\mbf{R}_k \cdot  d \mbf{r}_k \neq 0$  &
                  $\mbf{R}_k \cdot  \delta \mbf{r}_k \neq 0$ \\
\hfill {\scriptsize{$\mbf{R}_N = \mbf{T}_N$}} &
  \framebox[2.5cm]
  {{\scriptsize{$\displaystyle{\sum_{k=1}^{N}}$}} $\,\mbf{R}_k \cdot \delta \mbf{r}_k = 0$} &
   {\scriptsize{$\displaystyle{\sum_{k=1}^{N}}$}} $\,\mbf{R}_k \cdot d \mbf{r}_k \neq 0$ &
  \framebox[2.5cm]
  {{\scriptsize{$\displaystyle{\sum_{k=1}^{N}}$}} $\,\mbf{R}_k \cdot \delta \mbf{r}_k = 0$} \\
&&&\\
\hline
\end{tabular}

\begin{flushleft}
\hskip 1cm $^{\dagger}$ : $\delta \mbf{r}_k$ and $d \mbf{r}_k$ are {\it equivalent}
   if the constraints are both holonomic and scleronomous. \\
\hskip 1cm $^{\star}$ : $\delta \mbf{r}_k$ and $d \mbf{r}_k$ are {\it not equivalent}
   if the constraints are non-holonomic and/or rheonomous. 
\end{flushleft}
\end{table}

Let us now study the $N$-pendulum when its point of suspension is moving with 
velocity $\mbf{v}_0$. The system now has a rheonomous constraint as well,
\beq
|\mbf{r}_1 - \mbf{v}_0 t| = L_1, \hskip 1cm |\mbf{r}_k - \mbf{r}_{k-1}|=L_k,
\hskip 2cm k=2,3,\dots,N.
\eeq
In accordance with the previous section, for $N$-pendulum 
with moving support, the only non-trivial modification is that, 
the virtual displacements are no longer
equivalent with the allowed displacements. 
For the first bob $\mathcal{P}_1$,
\beqar\label{mv_ndisp1}
&& d \mbf{r}_1= \mbf{v}_{1} \, dt = \mbf{v}_{0} \, dt + dq_1 \hat{\mbf{n}}_1, \nonumber \\
&& \delta \mbf{r}_1= d \mbf{r}_1 - d \mbf{r}_1' 
=( \mbf{v}_{0} dt \hskip -.6cm \backslash \,\,\, \backslash \, + dq_1 \hat{\mbf{n}}_1 ) 
-( \mbf{v}_{0} dt \hskip -.6cm \backslash \,\,\, \backslash \, + dq_1' \hat{\mbf{n}}_1)
=( dq_1 -dq_1') \hat{\mbf{n}}_1
= \delta q_1 \hat{\mbf{n}}_{1}. 
\eeqar
Thus $ \delta \mbf{r}_1$ is a vector along $\hat{\mbf{n}}_1$, whereas 
$ d \mbf{r}_1$ is sum of a vector along  $\hat{\mbf{n}}_1$ and a vector 
along $\mbf{v}_{0}$.
For subsequent bobs $\mathcal{P}_k$,
\beqar\label{mv_ndisp2}
& d \mbf{r}_k &= \mbf{v}_k dt = \mbf{v}_{k-1} dt + b_k \hat{\mbf{n}}_k dt 
= \mbf{v}_0 dt + d q_1 \hat{\mbf{n}}_1 + d q_2 \hat{\mbf{n}}_2 
+ \dots + b_k \hat{\mbf{n}}_k dt \nonumber \\
& \delta \mbf{r}_k &= (\mbf{v}_0 dt \hskip -.6cm \backslash \,\,\, \backslash \,
     + d q_1 \hat{\mbf{n}}_1 + d q_2 \hat{\mbf{n}}_2 + \dots + d q_k \hat{\mbf{n}}_k) - 
     (\mbf{v}_0 dt \hskip -.6cm \backslash \,\,\, \backslash \,
     + d q_1' \hat{\mbf{n}}_1 + d q_2' \hat{\mbf{n}}_2 + \dots + d q_k' \hat{\mbf{n}}_k)
     \nonumber \\
&&= \delta q_1 \hat{\mbf{n}}_{1} +\delta q_2 \hat{\mbf{n}}_{2} 
+ \dots + \delta q_k \hat{\mbf{n}}_k.
\eeqar
Thus $\delta \mbf{r}_k$  and $d \mbf{r}_k$ are not necessarily equivalent.
However the relation between $ \delta \mbf{r}_{k-1}$ and $\delta \mbf{r}_k$
remains the same as in the previous case of $N$ pendulum with stationary support.
\beq
\delta \mbf{r}_k = \delta \mbf{r}_{k-1} + \delta q_k \hat{\mbf{n}}_{k}
\eeq
Hence the above inferences, in particular, (\ref{nw1}), (\ref{nw2}), (\ref{nwsum}) are 
true even for an $N$-pendulum with moving support.

\noindent
The adjacent table summarizes the results presented in sections II and III.

\section{Conclusion}
The zero virtual work principle of d'Alembert identifies a special class
of constraints, which is available in nature, and is solvable \cite{sj_sec23}.
Two noteworthy features of d'Alembert's principle are,
(i) it involves the {\it virtual work} $(\mbf{R}_k \cdot \delta \mbf{r}_k)$,
i.e., work done by constraint forces on {\it virtual displacement} 
$\delta \mbf{r}_k$ and not on allowed displacement $d \mbf{r}_k$, and
(ii) the total virtual work for the entire system vanishes, i.e.,  
$\sum (\mbf{R}_k \cdot \delta \mbf{r}_k)=0$, though virtual work on individual
particles of the system need not be zero $(\mbf{R}_k \cdot \delta \mbf{r}_k) \neq 0$.
For holonomic (velocity independent) and scleronomous (time independent) 
constraints, e.g., pendulum with stationary support, the allowed and virtual
displacements are collinear and hence a distinction between work done
on {\it allowed displacement} and that on {\it virtual displacement} is
not possible.
For understanding the nature of this distinction one needs to study
a system which is either non-holonomic or rheonomous.
A pendulum with moving support, or a particle sliding down a moving 
frictionless inclined plane are examples of simple rheonomous systems.
In constrained systems involving a single particle, the second feature 
mentioned above, in reference to d'Alembert's principle, becomes irrelevant.
As there is only one particle, there is no summation in virtual work,
and $(\mbf{R} \cdot \delta \mbf{r}) = 0$. This implies that the force of
constraint $\mbf{R}$ is normal to the virtual displacement $\delta \mbf{r}$.
In order to really appreciate the importance of summation in the {\it total
virtual work}, one needs to study system of particles involving several
constraints. 
The double pendulum and $N$-pendulum, particularly with moving support,
present two of the simplest systems illustrating the subtlety of 
d'Alembert's principle.


\begin{thebibliography}{99}
\bibitem{goldstein} H. Goldstein, {\it Classical Mechanics}, 
Addison-Wesley Publishing Co., Reading, Massachusetts, 1980.
\bibitem{greenwood} D. T. Greenwood, {\it Classical Dynamics}, 
Prentice Hall, New York, 1977.
\bibitem{hylleraas} E. A. Hylleraas, {\it Mathematical and Theoretical Physics},
vol. I, Wiley Interscience, New York, 1970. 
\bibitem{sommer} A. Sommerfeld, {\it Mechanics, Lectures on Theoretical Physics}, vol. I, Academic Press, New York, 1952.
\bibitem{taylor} T. T. Taylor, {\it Mechanics: Classical and Quantum}, 
Pergamon Press, Oxford, 1976.
\bibitem{sj_ejp_27} Ray S, Shamanna J 2006 Eur. J. Phys. {\bf 27} 311-329, physics/0510204.
\bibitem{sj_ejp_eqn} equations (10) and (14) of section 2 in Ray S, Shamanna J 2006 Eur. J. Phys. {\bf 27} 311-329.
\bibitem{sj_sec23} section 2.3 in Ray S, Shamanna J 2006 Eur. J. Phys. {\bf 27} 311-329.
\end{thebibliography}
\end{document}